\documentclass[prb,twocolumn,showpacs,groupedaddress,superscriptaddress]{revtex4-1}

\usepackage{bm,bbm,times,amsfonts,amsmath,amssymb,graphicx,graphics,xcolor,dcolumn}
\usepackage[next]{inputenc}
\usepackage[dvips]{epsfig}
\usepackage[pdfstartview=FitH]{hyperref}
\usepackage{natbib}

\def\be{\begin{equation}}
\def\ee{\end{equation}}

\def\bi{\begin{itemize}}
\def\ei{\end{itemize}}
\def\bn{\begin{enumerate}}
\def\en{\end{enumerate}}
\def\ba{\begin{array}}
\def\ea{\end{array}}
\def\la{\langle}
\def\ra{\rangle}

\def\ket#1{{\vert #1 \rangle}}

\begin{document}

\title{Quantum phase transitions in the Kondo-necklace model:\\
Perturbative continuous unitary transformation approach}

\author{S. Hemmatiyan}
\affiliation{Department of Physics, Texas A\&M University, College Station, TX 77843-4242, USA}
\affiliation{Department of Physics, Sharif University of Technology, Tehran 14588-89694, Iran}

\author{M. Rahimi Movassagh}
\affiliation{Department of Physics and Astronomy, McMaster University, Hamilton, ON L8S 4M1, Canada}
\affiliation{Department of Physics, Sharif University of Technology, Tehran 14588-89694, Iran}

\author{N. Ghassemi }
\affiliation{Department of Physics, Texas A\&M University, College Station, TX 77843-4242, USA}
\affiliation{Department of Physics, Sharif University of Technology, Tehran 14588-89694, Iran}

\author{M. Kargarian}
\affiliation{Department of Physics, Ohio State University, 191 West Woodruff Ave, Columbus, OH 43210, USA}

\author{A. T. Rezakhani}
\affiliation{Department of Physics, Sharif University of Technology, Tehran 14588-89694, Iran}

\author{A. Langari}
\affiliation{Department of Physics, Sharif University of Technology, Tehran 14588-89694, Iran}

\date{\today}

\begin{abstract}
The Kondo-necklace model can describe magnetic low-energy limit of strongly correlated heavy fermion materials. There exist multiple energy scales in this model corresponding to each phase of the system. Here, we study quantum phase transition between the Kondo-singlet phase and the antiferromagnetic long-range ordered phase, and show the effect of anisotropies in terms of quantum information properties and vanishing energy gap. We employ the ``perturbative continuous unitary transformations" approach to calculate the energy gap and spin-spin correlations for the model in the thermodynamic limit of one, two, and three spatial dimensions as well as for spin ladders. In particular, we show that the method, although being perturbative, can predict the expected quantum critical point, where the gap of low-energy spectrum vanishes, which is in good agreement with results of other numerical and Green's function analyses. In addition, we employ concurrence, a bipartite entanglement measure, to study the criticality of the model. Absence of singularities in the derivative of concurrence in two and three dimensions in the Kondo-necklace model shows that this model features multipartite entanglement. We also discuss crossover from the one-dimensional to the two-dimensional model via the ladder structure.
\end{abstract}


\maketitle

\section{Introduction}

Strongly correlated systems typically have various competing energy scales driving system into a variety of phases. At zero temperature, two different phases are separated by a quantum critical point, and essentially quantum fluctuations lead to a quantum phase transition (QPT),\cite{sachdev:book} where a macroscopic change occurs in physical properties of system. The role of correlations, either classical or quantum, between underlying degrees of freedom is a key to understanding QPTs. Entanglement is a measure of nonlocal correlations of quantum many-body states. Entanglement is a versatile quantity intensively studied in quantum information theory and strongly correlated systems providing an exchange of ideas between the two apparently distinct fields. Indeed, entanglement has been recognized as an essential resource in quantum computation and communication \cite{nielsen:book}. Ideas based on entanglement lead to distinguish criticality from off-criticality in one dimension \cite{calabrese:jpa04}, and conclude area law scaling of entanglement entropy \cite{eisert:rmp10} and topological sub-leading term \cite{levin:prl06,kitaev:prl06}.

The Kondo lattice model \cite{tsunetsugu:rmp97} is one of promising models in the study of strongly-correlated heavy fermion compounds \cite{hewson:book}, where partially filled shells (\emph{f}) are screened by outer shells (\emph{s} or
\emph{p}). Inner shells are localized, and can be effectively characterized by magnetic impurities located in the electron gas of
outer shells. In the presence of strong interaction between magnetic impurities and electron gas, a more simplified version of the Kondo lattice, namely the Kondo-necklace (KN) model \cite{doniach:77} arises. The phases of this model in the spatial dimensions $d=1$, $2$, and $3$ have been studied by different methods. While Monte Carlo \cite{scalettar:prb85} and density-matrix renormalization group \cite{moukouri:prb95} (DMRG) results confirm the absence of QPT in one dimension, a phase transition between the antiferromagnetic and Kondo-singlet phases occurs in higher dimensions.

Here we study QPT(s) of the anisotropic KN model in the spatial dimensions $d = 1$, $2$, and $3$ by employing the perturbative continuous unitary transformations (PCUT) method \cite{wegner:ann94,wilson1,wilson2,knetter:EJB00}. The extreme limit of the model, where the interactions between dimers are suppressed, is characterized as isolated dimers, i.e., a Kondo-singlet state. The ground state is a liquid of singlet states and excited states are equally separated. Due to the latter point, the model is well suited to the PCUT method, where the inter-dimer interaction is treated perturbatively. The method also allows us to study the system at the thermodynamic limit and no finite-size scaling effect is needed. Moreover, in contrast to other numerical methods, we can obtain energy of states analytically as a function of Hamiltonian parameters. It helps us find some other underlying phase transition witnesses such as concurrence (from correlation functions that could be calculated from derivatives of the ground-state energy with respect to Hamiltonian parameters). In addition to the above advantages, we do not need to \textit{a priori} know  eigenstates of the system in order to find energy spectrum of the system---the energies are calculated independent of finding the eigenstates of the system. And, in this method, fortunately there is no dimensional restriction (which obstructs some other numerical methods). Indeed, we can use this method for arbitrary dimensions and system sizes.

We further characterize the QPT by using ``concurrence" \cite{wooter:prl98} as a measure of entanglement between a pair of spin-$1/2$ particles. The entanglement properties of the one-dimensional KN model with a few sites have already been addressed in the literature \cite{Saguia:pra03}. In the current work, we consider the thermodynamic limit via PCUT and probe phase transitions in spatial dimensions $d=1$, $2$, and $3$. Despite its local nature, nonanalytic behavior of derivatives of concurrence signals a phase transition in the system, and its scaling in the vicinity of critical points is connected to the universality class of the model \cite{Osterloh:na02,kargarian07,Amico2009}. We will show that concurrence does not capture the critical properties of the KN model at the mentioned spatial dimensions, manifesting that the underlying correlations are not of the bipartite nature. Although our model has continuous symmetry, the absence of bipartite correlations is similar to the DMRG results~\cite{mendoza2010} for the one-dimensional model with discrete symmetry. However, the critical points detected by the gap closing points are in good agreement with the mean-field theory analysis \cite{Langari06,Thalmeier07,mahmoudian} and the Green's function approach \cite{rezania}. 

Moreover, we discuss crossover from the one-dimensional model to the critical two-dimensional model via the ladder geometry. Despite the spin-$1/2$ antiferromagnetic Heisenberg model which reveals even/odd universality classes for the $n$-leg ladder \cite{Dagotto1996}, we observe that seemingly an $n$-leg KN model belongs to the two-dimensional universality class. Our conclusion is based on the $4$th order PCUT approach, which necessitates more investigations on different aspects of the crossover between one and two spatial dimensions.

This article is organized as follows. In Sec.~\ref{pcut}, the PCUT method is briefly reviewed. In Sec.~\ref{pcut_kn}, the explicit expression for the ground-state energy and excitation spectrum of the KN model are derived by the PCUT method, which are next used to study the QPT of the model in Sec.~\ref{qpt}. The article is concluded by a summary and discussion of our results.

\section{Perturbative continuous unitary transformation (PCUT)}
\label{pcut}

The number of examples of strongly correlated systems with exact solutions (exact ground state, excitation spectrum, and correlation functions) is rather rare, thus one often needs to resort to numerical methods to obtain (an approximation of) the physical properties of these systems. One of these exact methods---introduced independently in the contexts of condensed matter and quantum chromodynamics problems---to diagonalize a Hamiltonian properly, regardless of system size, is the continuous unitary transformation \cite{wegner:ann94,wilson1,wilson2}. In this approach, the Hamiltonian is considered as a function of a flow parameter $\ell$. The Hamiltonian is transformed to a simpler form (diagonal or band-block diagonal) under a flow equation, which is based on applying infinite numbers of infinitesimal unitary transformation.

Let us define the Hamiltonian $H(\ell)$ as a function of a continuous parameter $\ell$, where $H(0)$ is the bare Hamiltonian and $H(\infty)$ is the final (block-) diagonalized effective Hamiltonian. The evolution of the Hamiltonian in terms of the continuous unitary transformation is given by $H(\ell)=U(\ell)H(0)U^{\dag}(\ell)$, or equivalently by 
\begin{align}
\frac{dH(\ell)}{d\ell}=[\eta(\ell),H(\ell)],
\label{hofl}
\end{align}
where $U(\ell)$ is a unitary transformation, and the anti-Hermitian operator $\eta(\ell)= (dU/d\ell)U^{\dag}$ is the generator of this transformation. There are several choices for $\eta(\ell)$ which depend on the Hamiltonian and the possibility to find a closed form for the transformation. For instance, Wegner \cite{wegner:ann94} considered $H(\ell)=H_{\mathrm{d}}(\ell)+H_{\mathrm{nd}}(\ell)$ where $H_{\mathrm{d}}$ is the diagonal part of Hamiltonian in a specified basis, and $H_{\mathrm{nd}}$ is the corresponding off-diagonal part, and suggested 
\begin{equation}
\eta(\ell)=[H_{\mathrm{d}}(0),H(\ell)].
\end{equation}
However, this choice does not lead to an analytic solution for all models when $\ell\rightarrow \infty$. Moreover, Wegner's choice mixes the blocks of the Hamiltonian during the transformation if the original Hamiltonian is block-diagonal. To overcome this problem, Mielke \cite{Mielke:epjb98} proposed a generator to preserve the band-diagonal form of the initial Hamiltonian, and accordingly Knetter and Uhrig introduced PCUT \cite{knetter:EJB00}. In this approach, the original Hamiltonian is decomposed
to two parts $H(0)=H_0+\lambda H_s$, where the unperturbed Hamiltonian is $H_0=H_{\mathrm{d}}(0)$, and $H_s$ represents the off-diagonal part at $\ell=0$ which is considered as a perturbation for $H_0$. The generator of transformation is obtained order by order in terms of the perturbation parameter $\lambda$. The off-diagonal part evolves to reach the $\ell=\infty$ effective Hamiltonian.

PCUT has two prerequisites: i. the unperturbed Hamiltonian ($H_0$) needs to be composed of equidistant energy levels which are bounded from below; ii. the perturbing Hamiltonian ($H_{s}$) should be written as $H_s=\sum_{n=-N}^{N} T_{n}$ (for $N<\infty)$, where $T_n$ increases or decreases the energy quanta of the specified levels of $H_0$, i.e., $[H_0, T_n]=nT_n$.

We note that the spectrum of $H_0$ can be labeled with the eigenvalues of the quasiparticle number operator $Q$. For example, 
in the KN model (see below), the ground state is the product of singlet states, and excitations are local triplets on each site. Hence, the number of the triplets is given by the quasiparticle operator $Q$. Clearly, the $T_n$ operators with $n\neq 0$ connect distinct sectors with different quasiparticle number, and $T_0$ spreads quasiparticle within each sector. The original Hamiltonian $H$ can be replaced with an effective one $H_{\mathrm{eff}}$ which conserves the number of quasiparticles. The Hamiltonian $H_{\mathrm{eff}}$ is obtained by continuous unitary transformation; that is, $H_{\mathrm{eff}}=H(\infty)$ (satisfying $[H(\infty), Q]=0$) \cite{knetter:EJB00,knetter:JPA03}. Because of the conservation relation, one can rewrite 
\begin{equation} 
H_{\mathrm{eff}}=\sum_{q\in \mathbb{N}} H_{\mathrm{eff}|q},
\label{H_{eff}}
\end{equation}
where $H_{\mathrm{eff}|q}$ denotes the restricted operator acting on a sub-space of the Hilbert space spanned by the $q$-quasiparticles. Therefore we can study each sector defined by the number of quasiparticles.
 
In order to obtain $H_{\mathrm{eff}}$, we follow the formalism developed by Uhrig and Knetter,\cite{knetter:EJB00}, where the generator of the transformation is defined by
\begin{align}
\eta(\ell)=\sum_{k=1}^{\infty} \lambda^{k} \sum_{\vert\underline{m}\vert =k }
\mathrm{sgn}[M(\underline{m})]F(\ell; \underline{m})T(\underline{m}),
\label{2}
\end{align}
where $k$ is the order of perturbation, $F(\ell;\underline{m})$ are real-valued functions obtained by some recursive differential
equations \cite{knetter:EJB00}, $M(\underline{m})=\sum _{i=1} ^{k} m_{i}$, and $\underline{m}$ denotes a sequence of quanta labels
\begin{align}
\underline{m} \equiv (m_{1}, m_{2}, m_{3},\ldots, m_{k}), \nonumber\\
m_{i} &\in&\lbrace 0, \pm 1, \pm2, \dots \pm N\rbrace, 
\end{align}
and 
\begin{align}
T(\underline{m})\equiv T_{m_1}T_{m_2}T_{m_3} \dots T_{m_k}.
\label{126}
\end{align}
Note that $\vert \underline{m}\vert =k$ implies the summation over all possible configurations of a sequence with $k$ members, and $M(\underline{m})$ is the total number of energy quanta created or annihilated by $T(\underline{m})$. Thus the Hamiltonian is given in the following form:
\begin{equation}
\label{hl}
H({\ell})=H_{0}+\sum_{k} \lambda^{k} \sum _{\vert \underline{m}\vert =k}
F(\ell;\underline{m}) T(\underline{m}).
\end{equation}

We remind that $\eta(\ell)$ of Eq. (\ref{2}) keeps only those processes that conserve the number of quasiparticles and eliminates all parts of $H$ changing the number of quasiparticles\cite{knetter:EJB00,knetter:JPA03}.

To obtain a final diagonalized Hamiltonian, the number of energy quanta created or annihilated at $\ell \to \infty$ must be zero.
Therefore, we can write the effective Hamiltonian as
\begin{equation}
H_{\mathrm{eff}}=\lim_{\ell\to\infty} H(\ell) =H_{0}+\sum_{k=1}^\infty \lambda^k \sum_{| \underline{m} | = k,
M(\underline{m})=0} C(\underline{m}) T(\underline{m}),
\label{31b}
\end{equation}
where $C(\underline{m})\equiv F(\infty; \underline{m})$. The solution of the flow equation \eqref{hofl} for the generator given by Eq.~\eqref{2} and the Hamiltonian defined in Eq.~\eqref{hl} leads to the coefficients $C(\underline{m})$. Hence, the effective Hamiltonian is obtained perturbatively as a sum of different orders of perturbation.

\section{PCUT for the Kondo-necklace (KN) model}
\label{pcut_kn}

\begin{figure}
\includegraphics[width=6cm]{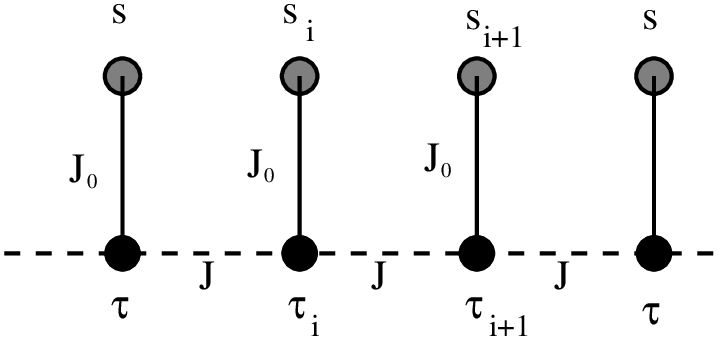}
\caption{One-dimensional KN model. Solid and gray circles denote, respectively, spins of itinerant and localized electrons.}
\label{kn-chain} 
\end{figure}

The anisotropic KN model on a hypercubic lattice is defined by the following Hamiltonian:
\begin{align}
H  = & J_0 \sum_{i=1
}^{L} \big(s_{i}^{x}\tau_{i}^{x}+s_{i}^{y}\tau_{i}^{y}+\Delta s_{i}^{z}\tau_{i}^{z}\big) \nonumber\\
&+ J \sum_{\langle i,j\rangle} \big(\eta_{x} \tau _{i} ^{x}\tau_{j} ^{x} +
\eta _{y} \tau _{i} ^{y}\tau_{j} ^{y}+\eta _{z} \tau _{i} ^{z}\tau_{j} ^{z}\big),
\label{Hamiltonian}
\end{align}
where $\tau_i$ is the spin of itinerant electrons on the $d$-dimensional hypercubic lattice with $L^d$ sites, $s_i$ is the spin of localized electrons connected to each lattice point, $\langle i,j\rangle$ represents the nearest neighbor sites on the lattice, and $J_0$ and $\Delta$ are, respectively, the local exchange coupling and anisotropy parameters. Here $J$ and $\eta_{\mu}$ are the exchange and anisotropy couplings on the lattice, respectively. The one-dimensional lattice is shown in Fig.~\ref{kn-chain}, from whence the two- and three-dimensional cases can also be easily understood (see Fig. \ref{ladder}). We have considered the fully anisotropic case, which enables us to study the anisotropy effects as well to obtain the correlation functions via introducing a generating function. For simplicity, we define $\lambda=J/J_0$ and set $J_0=1$.

To apply PCUT, we consider the following decomposition of the Hamiltonian:
\begin{equation}
H =H_{0}+ \lambda H_{s}, 
\label{c11}
\end{equation}
where
\begin{align}
H_{0} &= \sum_{i} \mathbf{s}_i \cdot \bm{\tau}_{i}, \\
H_{s}=\alpha \sum_{i} s_{i}^z \tau_{i}^z &+\sum_{\langle i,j\rangle} (\eta_{x} \tau_{i}^{x}\tau_{j}^{x}+\eta_{y} \tau_{i}^{y}\tau_{j}^{y}+ \eta_{z} \tau_{i}^{z}\tau_{j}^{z}), 
\end{align}
in which $\alpha=(\Delta-1)/\lambda$. The anisotropic part of the local interactions are considered in the perturbing part $H_s$ in order to have an equidistant spectrum for $H_{0}$ (the first prerequisite of PCUT). The ground state of $H$ is a Kondo-singlet state for $J_0 \gg J$ which justifies $\lambda$ to be small. In other words, we trace the quantum phase transition from the Kondo-singlet phase for which $\lambda$ is a small parameter. The Kondo-singlet state is the ground state of the unperturbed Hamiltonian $H_{0}$, which is a product of singlet states on local dimers ($\tau_i$-$s_i$ pair).

Moreover, the spectrum of a local dimer is composed of a singlet and triplet state representing two quanta, $0$ and $1$, respectively. Thus the effect of $H_s$ on two neighboring singlet states changes the quanta by $n\in\{0,1,2\}$, which provides the same coefficients $C(\underline{m})$ derived in Ref.~\onlinecite{knetter:EJB00} for our calculations, too.

\subsection{$T_{n}$ operators}

To obtain $H_{\mathrm{eff}}$, we need to calculate the $T(\underline{m})$ operators which essentially define the effect of $H_s$ on a base ket. In the case that the Hamiltonian is composed of two-body interactions, the effect of $H_s$ can be represented by its effect on two neighboring dimers, i.e.,
\begin{equation}\label{hs}
H_s=\sum_{\langle i,j\rangle} H^{i,j}_{s}.
\end{equation}
A dimer is a pair of $\tau_i$ spin and its corresponding neighboring impurity spin $s_i$, which can be in a singlet or triplet state configurations. Accordingly, the effect of $H^{i,j}_{s}$ on two neighboring dimers $\langle i,j\rangle$ may change the quantum number of the state by $n=0, \pm1, \pm2$, which is represented by $q_{n}$ (see Appendix \ref{aa}). Therefore, the pair Hamiltonian of $H_s$ is written in the following form:
\begin{align}
H^{i,j}_{s} =&\alpha (\tau^{z}_{i} s^{z}_{i}+\tau^{z}_{j} s^{z}_{j}) +\eta_{x} \tau^{x}_{i} \tau^{x}_{j}+\eta_{y} \tau^{y}_{i} \tau^{y}_{j}
+ \eta_{z} \tau^{z}_{i} \tau^{z}_{j} \nonumber \\
&=q_{-2}+q_{-1}+q_{0}+q_{1}+q_{2}.
\label{c14}
\end{align}
Two neighboring dimers can take $16$ configurations depending on that the configuration of each dimer is a singlet or triplet state (see Table \ref{jadv} of Appendix \ref{aa}). The translational symmetry of the Hamiltonian allows to use Table \ref{jadv} for any pair of neighboring dimers. According to Eq.~\eqref{hs}, the perturbing Hamiltonian $H_s$ is written in terms of a sum over the $q_{n}$ operators. For example, in the one-dimensional model consisting of $L$ dimers with periodic boundary condition, we 
have
\begin{align}
\label{c19}
\sum^{L}_{i=1} H^{i,i+1}_{s} &= \sum^{L}_{i=1} q_{-2} + \sum^{L}_{i=1} q_{-1} + \sum^{L}_{i=1} q_{0} + \sum^{L}_{i=1} q_{1} + \sum^{L}_{i=1} q_{2} \nonumber \\
&=T_{-2} + T_{-1} +T_{0} +T_{1} +T_{2}.
\end{align}
Here $T_n$ is the sum of $q_{n}$ operators on all bonds which connects two neighboring dimers on the lattice. One can also see that some of the $C(\underline{m})$ coefficients in the effective Hamiltonian vanish for some subtle reasons. 

\begin{table}[tp]
\caption{The GSE per site for the one-dimensional KN model with $(\eta_x,\eta_y,\eta_z)=(1,1,0)$ and $\Delta=1$. 
A comparison between the $4$th order PCUT and Lanczos results (up to 8-digits of accuracy) is given by the 
relative error $\vert \epsilon_{\mathrm{Lanczos}} -\epsilon_{\mathrm{PCUT}} \vert/\epsilon_{\mathrm{Lanczos}}$.}
\begin{tabular}{|c|c|c|c|}
\hline\hline	
 $\lambda$&$\epsilon _{\mathrm{Lanczos}}$&$\epsilon _{\mathrm{PCUT}}$& relative error \\ \hline
\hline $0$&   $-0.37500000$&$-0.37500000 $&$0$\\
\hline $0.1$&$-0.37531250$&$-0.37531250$&$0.00000000$\\
\hline $0.2$&$-0.37624999$&$-0.37625000$&$0.00000003$\\
\hline  $0.3$&$-0.37777517$&$-0.37781250$&$0.0001$\\
\hline$0.4$&$-0.37999946$&$-0.38000000$&$0.000001$\\
\hline$0.5$&$-0.38281007$&$-0.38281250$&$0.000006$\\
\hline$0.6$&$-0.38631563$&$-0.38625000$&$0.0002$\\
\hline$0.7$&$-0.39040284$&$-0.39031250$&$0.0002$\\
\hline$0.8$&$-0.39496177$&$-0.39500000$&$0.0001$\\
\hline$0.9$&$-0.40020219$&$-0.40031250$&$0.0003$\\
\hline$1$& $-0.40611293$&$-0.40625000$&$0.0003$\\
\hline\hline
\end{tabular}
\label{comparison}
\end{table}

\subsection{Ground-state energy}

According to PCUT, the ground-state energy (GSE) is given by $\la0\vert H_{\mathrm{eff}} \vert0\ra$, where $H_{\mathrm{eff}}$ is given in Eq.~(\ref{31b}), and $\ket{0}$ is the direct product of singlet states over all dimers. We have calculated the GSE per spin ($\varepsilon= E_0/(2L^{d}$)) for the fully anisotropic model and on $d$-dimensional hypercubic lattice up to the $4$th order of perturbation. Note that there are two spins ($\tau$, $s$) corresponding to each lattice point, $\tau$ on the hypercubic lattice and $s$ a local spin. Moreover, the following result is valid for a lattice in the thermodynamic limit ($L\rightarrow \infty$):
\begin{align}
\varepsilon =&- \frac{3}{8} - \frac{(2\alpha d)}{2}  \overline{ \lambda}
	  -\frac{d\overline{ \lambda} ^{2}}{4} ( \eta_{x} ^{2}+  \eta_{y} ^{2}+  \eta_{z} ^{2} )  \nonumber\\
	&+\frac{d}{4}\overline{\lambda}^{3} \Big(-3\eta_{x}\eta_{y}\eta_{z} +2 (2\alpha d) (\eta_{x}^{2}+\eta_{y}^{2}) \Big) \nonumber\\
	  & + \frac{d}{16} \overline{\lambda}^{4} \Big( (4d-1) ( \eta_{x} ^{2}+  \eta_{y} ^{2}+
\eta_{z} ^{2} ) ^{2} -4\eta_{z}^{2}( \eta_{x}^{2} +\eta_{y}^{2})\nonumber\\
&-4\eta_{x}^{2}\eta_{y}^{2}-4(4d-3) ( \eta_{x} ^{4}+
\eta_{y} ^{4}+  \eta_{z} ^{4})\nonumber\\
&-16(2\alpha d)^{2}( \eta_{x}^{2} +\eta_{y}^{2})+32(2\alpha d) \eta_{x}\eta_{y}\eta_{z}\Big),
\label{gse-perspin}
\end{align}
where $\overline{\lambda}= \lambda / 4$. The first term in the above equation is the singlet GSE (i.e., the zeroth-order value). 
We have compared the PCUT results with the results of the Lanczos diagonalization for the one-dimensional isotropic case in Table \ref{comparison}. The Lanczos computations have been done on several lengths up to $24$ spins ($L=6, 8, 10, 12$), where we found a small finite-size dependence. This indicates that the model with $24$ spins is already large enough to be exhibiting a reliable approximation of the thermodynamic limit. The last column in Table \ref{comparison} shows the error of the PCUT results compared with the Lanczos results, where the maximum error is less than $0.1 \%$ for $\lambda=1$.
It confirms that the PCUT results are reliable as long as $\lambda<1$. We have also made a similar comparison for other values of anisotropies, which essentially gives the same conclusion.

\begin{figure}[tp]
\includegraphics[width=6cm]{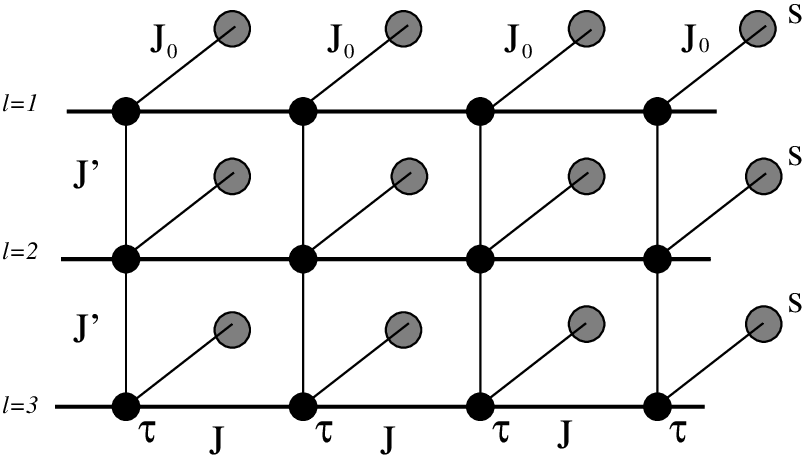} 
\caption{Three-leg KN ladder. Solid and gray circles denote, respectively, spins of itinerant and localized electrons.}
\label{ladder}
\end{figure}

We have also calculated the GSE of the KN model on a ladder, which is a quasi one-dimensional model. A three-leg ladder is shown in Fig. \ref{ladder}. This is important to study the crossover between one- and two-dimensional systems. The $n$-leg KN ladder is a set of $n$ KN chains which interact via the $\tau$-spins as given by the following Hamiltonian:
\begin{align}
H=&J_0 \sum_{i=1}^{L}\sum_{j=1}^{n} \mathbf{s}_{ij}\cdot \mathbf{\tau}_{ij} \\
&+ J 
\sum_{i=1}^{L+1}\sum_{j=1}^{n} (\eta_{x} \tau _{ij} ^{x}\tau_{i+1,j} ^{x}+
\eta_{y} \tau _{ij} ^{y}\tau_{i+1,j} ^{y}+\eta _{z} \tau_{ij} ^{z}\tau_{i+1,j} ^{z})
\label{301}\nonumber\\
&+J'
\sum_{i=1}^{L}\sum_{j=1}^{n-1}(\eta^{'}_{x} \tau _{ij} ^{x}\tau_{i,j+1} ^{x}+
\eta^{'}_{y} \tau _{ij} ^{y}\tau_{i,j+1} ^{y}+\eta^{'}_{z} \tau _{ij} ^{z}\tau_{i,j+1} ^{z}),
\nonumber
\end{align}
where the local term is isotropic, the anisotropy parameters along the chain direction are given by $\eta_{\alpha}$, and the anisotropy along the rung direction is represented by $\eta'_{\alpha}$. The exchange interaction along the rungs is $J'$, and we define $x=J'/J$. The periodic boundary condition is assumed along the leg direction, while the boundary condition is open along the rungs. The GSE per spin for an $n$-leg KN ladder to the $4$th order of the perturbation expansion in the PCUT approach is obtained 
\begin{widetext}
\begin{eqnarray}
\label{ladder-energy}
\varepsilon^{\mathrm{ladder}} &=& \frac{E^{\mathrm{ladder}}_{0}}{2 n L}=-\frac{3}{8}
-\frac{\overline{\lambda}^{2}}{4} \Big[( \eta_{x} ^{2}+  \eta_{y} ^{2}+  \eta_{z} ^{2} )
+x^{2}\frac{(n-1)}{n}( \eta ^{'2}_{x}+\eta ^{'2}_{y}+  \eta ^{'2}_{z} )\Big]
-\frac{3\overline{\lambda}^{3}}{4}\Big[\eta_{x} \eta_{y} \eta_{z}
+\frac{(n-1)}{n}\eta^{'}_{x} \eta^{'}_{y} \eta^{'}_{z}\Big] \nonumber\\
&& +\frac{\overline{\lambda}^{4}}{16}
\Big[8x^{2}\frac{(n-1)}{n}( \eta_{x} ^{2}+  \eta_{y} ^{2}+  \eta_{z} ^{2})
( \eta ^{'2}_{x}+  \eta ^{'2}_{y}+  \eta ^{'2}_{z})
+3x^{4}\frac{(n-1)}{n}( \eta ^{'2}_{x}+  \eta ^{'2}_{y}
+  \eta ^{'2}_{z})^{2}+2(\eta_{x}\eta_{y})^{2}+2(\eta_{x}\eta_{z})^{2} \nonumber\\
&& +2(\eta_{y}\eta_{z})^{2}
-\eta_{x}^{4}-\eta_{y}^{4}-\eta_{z}^{4}
-\frac{(n-1)}{n}(32 x^{2}+4x^4)(\eta ^{'4}_{x}+\eta ^{'4}_{y}+\eta ^{'4}_{z})
-4x^{4}\frac{(n-1)}{n}((\eta_{x}^{'}\eta ^{'}_{y})^{2}+(\eta'_{x}\eta'_{z})^{2}
+(\eta'_{y}\eta ^{'}_{z})^{2})\Big] .\nonumber\\
\end{eqnarray}
\end{widetext}
The above equation reduces to the energy of the one-dimensional KN model for $n=1$ and also for $x=0$. It should be noted that Eq.~(\ref{ladder-energy}) has been calculated for the isotropic local interaction $\alpha=0$.

\subsection{Excitation spectrum}

The one-magnon dispersion can be obtained using PCUT. The lowest excited energy is created by exciting a singlet dimer to a triplet in the ground state of Kondo-singlet. Let us represent the one-triplet state by $\vert j \rangle = | s,s, \cdots ,t_j, \cdots, s \rangle $, where the singlet dimer at position $j$ has been replaced by a triplet state. Among the three states in the triplet set ($t^{\pm}, t^0$), the energy of $t^{\pm}$ is lower than $t^0$. Thus we choose either $t^+$ or $t^-$ for constructing the one-triplet state. By virtue of $[H_{\mathrm{eff}}, Q]=0$, the effective Hamiltonian conserves the number of triplets. Indeed, in the one-particle sector the local excitation (triplet) propagates through the lattice. The one-triplet states $\vert j \rangle$ are used to build a basis for the one-particle sector. Since the Hamiltonian conserves the number of triplets, all states with a single triplet have the same energy. Hence, the linear combination of all one-triplet states at different positions is an elementary excited state represented by $|\mathbf{k}\rangle$ (a magnon for a $d$-dimensional hypercubic periodic lattice)
\begin{equation}
|\mathbf{k}\rangle = \frac{1}{\sqrt{L^d}} \sum_{\mathbf{r}} e^{i\mathbf{k}.\mathbf{r}} |ss \cdots s,t^{+}_{\mathbf{r}},s \cdots ss\rangle.
\label{k-state}
\end{equation}

\begin{figure}[tp]
\includegraphics[width=8cm]{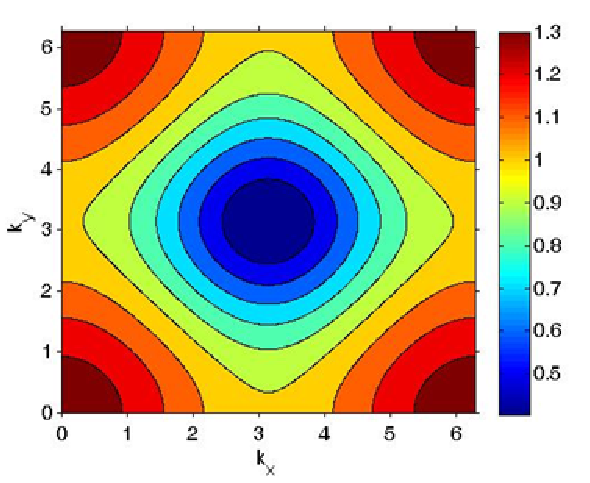}
\caption{Contour plot of the gap function $G_2(k_x,k_y)$ of the two-dimensional KN model. The gap obtains its minimum at ($\pi,\pi$). The corresponding parameters are $\lambda=0.5$, $(\eta_x,\eta_y,\eta_z)=(1,1,0)$, and $\alpha=0$.}
\label{cont1} 
\end{figure}

The dispersion of the magnon spectrum is calculated by the following equation:
\begin{equation}
\omega(\mathbf{k}) = \langle \mathbf{k}| H_{\mathrm{eff}} |\mathbf{k} \rangle - E_{0}.
\label{c27}
\end{equation}
Due to the existence of a triplet dimer in the excited state, the calculation of Eq. (\ref{c27}) increases dramatically for the $4$th order of perturbation. Thus we restricted our calculation to the $3$rd order of perturbation for the one-magnon spectrum. The magnon dispersion of the one-dimensional model is expressed by the following relation (with $\eta_x=\eta_y=1$):
\begin{widetext}
\begin{align}
\omega_1(k) =&1+\overline{\lambda}[4\alpha + 2 \cos(k_x)]-{\overline{\lambda}}^2[{\eta_{z}}^2-4 \eta_{z} \cos(k_x)+\cos(2k_x)] \nonumber \\
&+ \overline{\lambda}^{3}[-\frac{17}{2}\alpha+4 \alpha \cos(2k_x)-5 \cos(k_x)+\cos(3k_x)+\frac{13}{4}\eta_{z}+2 \eta_{z}\cos(2k_x)-\frac{7}{2}{\eta_z}^2\cos(k_x)].
\label{1d-dispersion}
\end{align}
\end{widetext}
Note that in contrast to the GSE, the spatial dimension does not enter directly into the dispersion of magnons, because the calculation depends on the position of the ground state triplet (excited dimer). A similar calculation leads to the magnon dispersion for two- and three-dimensional hypercubic lattice---see Appendix \ref{spectrum}.

The minimum of the magnon energy defines the energy gap, which appears at the antiferromagnetic wave vector $k_x=\pi$ for the one-dimensional model and gives
\begin{align}
G_1 =&\frac{1}{2}+\frac{\Delta}{2}+\overline{\lambda}[-2]+{\overline{\lambda}}^2[\frac{25}{16}-\frac{9}{16}\Delta+4 \eta_z+{\eta_z}^2] \nonumber \\
&+{\overline{\lambda}}^3[4+\frac{21}{4}\eta_z+\frac{7}{2}{\eta_z}^2].
\label{g1}
\end{align}
The gap $G_1$ is always nonzero for the interested range of the parameters ($\lambda<1$), thus representing no magnetic order for the one-dimensional model, whereas $G_2$ and $G_3$ behave differently. The magnon dispersion of the two-dimensional lattice is plotted in Fig. \ref{cont1} for $\lambda=0.5$, $(\eta_x,\eta_y,\eta_z)=(1,1,0)$, and $\alpha=0$. The excitation energy obtains its minimum at $\mathbf{k}=(\pi, \pi)$, which verifies the antiferromagnetic ordering.

For the two-dimensional lattice, the energy gap is given by
\begin{equation}
G_2=\Delta-4\overline{\lambda}-{\overline{\lambda}}^2[8 \eta_z+ 2 \eta_z^2+10 \Delta-4] +{\overline{\lambda}}^3[21 \eta_z^2+40 \eta_{z}-1].
\label{g2}
\end{equation}
Similar calculations show that for three-dimensional lattice the excitation energy is minimized at the antiferromagnetic vector $\mathbf{k}=(\pi, \pi, \pi)$, where the energy gap is
\begin{align}
G_3=&-\frac{1}{2}+\frac{3}{2} \Delta- 6 \overline{\lambda}
+{\overline{\lambda}}^2(-\frac{57}{2}-12\eta_z+ 3 \eta_z^2
+\frac{27}{2}\Delta)\\
&+{\overline{\lambda}}^3(\frac{135}{2}- 69\eta_z+\frac{105}{2}\eta_z^2).
\label{g3}
\end{align}

The analysis of $G_2$ and $G_3$ are given in the next section where the associated quantum critical point is recognized at the position of vanishing gap. Some caution should be taken about the accuracy and convergence of the gap by increasing orders of perturbation. The smaller parameters $\lambda$ and $\alpha$ in the perturbed Hamiltonian in Eq. (\ref{c11}), the more accurate results would be. For example, for gap functions the limit $\lambda \rightarrow 0$ and $\Delta \rightarrow 1$ should be taken simultaneously; otherwise, the perturbation expansion breaks down.  

\section{Quantum phase transition}
\label{qpt}

The QPT in the KN model is a competition between the Kondo-singlet phase and the antiferromagnetic long range order. The $U(1)$-symmetric one-dimensional model is always in the Kondo-singlet phase for any value of $\lambda$ \cite{gu,Langari06}, 
which agrees with the nonzero gap $G_1$ obtained in Eq.~\eqref{g1}. However, the $Z_2$-symmetric one-dimensional model 
exhibits a QPT to the antiferromagnetic order at $\lambda_c=2.22$ for the Ising interaction between
$\tau$-spins, i.e., $\eta_x=\eta_y=0$ and $\eta_z=1$ \cite{mahmoudian}. 

Unlike the one-dimensional model, both two- and three-dimensional KN Hamiltonian with $U(1)$ symmetry show a QPT from the disordered Kondo-singlet to the antiferromagnetic ordered phase \cite{gu,Langari06,rezania}. The quantum critical point depends on the anisotropy parameters; however, $\lambda_c<1$ for both cases. This observation supports our approach to study the QPTs in the KN model by PCUT.

We study the QPT of the KN model by using two different criteria: i. the closure of gap (where the gap vanishes at a quantum critical point), ii. the derivative of bipartite entanglement (which could show a singular behavior at a quantum critical point). It should be re-emphasized that PCUT gives results for the thermodynamic limit ($L\rightarrow \infty$), although its accuracy is given by the order of perturbation.

\subsection{Energy gap}

\begin{figure}
\includegraphics[width=8cm]{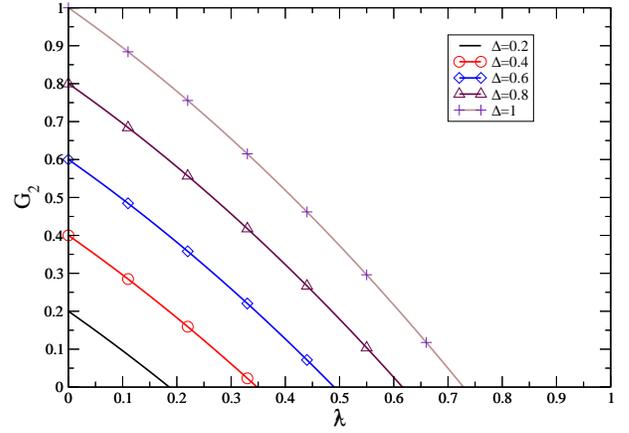}
\caption{The energy gap $G_2$ vs. $\lambda$ for different values of $\Delta$, and for $(\eta_x,\eta_y,\eta_z)=(1,1,0)$ in the two-dimensional lattice.}
\label{gap2d}
\end{figure}

\begin{figure}[tp]
\includegraphics[width=8cm]{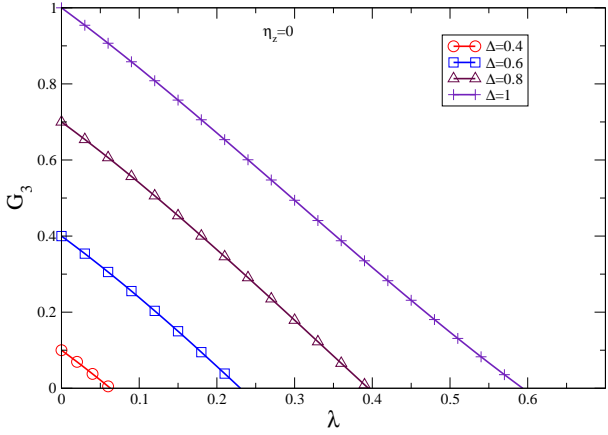}
\caption{Energy gap $G_3$ vs. $\lambda$ for different values of $\Delta$ and $\eta_x=\eta_y=1$, $\eta_z=0$ in the three-dimensional lattice.}
\label{gap3d} 
\end{figure}

We obtain the quantum critical point of the two- and three-dimensional KN model for different values of the anisotropy parameters. Figures \ref{gap2d} and \ref{gap3d} depict the energy gap of the two- and three-dimensional models [Eqs. \eqref{g2} and \eqref{g3}], respectively. Different plots belong to different anisotropy values ($\Delta$). This dependence instead implies dependence of the quantum critical point on $\Delta$. For the two-dimensional model, the results from other methods such as the mean-filed \cite{Langari06} and Green's function \cite{rezania} approaches are also available. Table \ref{qcp2d} summarizes this $\Delta$-dependence of the mentioned methods. The results of all methods are in agreement with each other (except at small values of $\Delta$). It is seen that as the anisotropy parameter $\Delta$ on the bonds increases, the phase transition at the critical point $\lambda_c$ occurs at higher values. This is plausible because the gap between singlet and triplet states increases with $\Delta$. Thus a higher $\lambda_c$ is needed to close the gap and drive the Kondo-singlet phase to the antiferromagnetic phase.

\subsection{Concurrence}

The role of entanglement in characterizing QPT has been exhaustively studied in the past few years. In particular, in Ref. 
\onlinecite{Osterloh:na02} it has been shown that ``concurrence" (as a measure of bipartite entanglement) \cite{wooter:prl98}---although a local quantity---can signal QPT of some physical models. Specifically, it has been illustrated that the derivative of concurrence with respect to the control parameter diverges at the quantum critical point of the Ising model in a transverse magnetic field. Such divergent behavior is a signature of QPT, while it does not reveal directly the type of ordering beyond the critical point. In this example, finite-size scaling and critical exponents coincide with the universality class of the associated phase transition. The nonanalytic behavior of quantum correlations (encapsulated in entanglement) has been discussed for spin models and itinerant systems in several previous works \cite{Wu,Vidal1,Vidal2,Bose,Osborne,Verstraete,Zanardi,Gu,Anfossi}.

Concurrence of a pair of spin-$1/2$ is obtained by the following expression:
\begin{align}
C=\max \lbrace 0, \sqrt{\lambda_{1}}-\sqrt{\lambda_{2}}-\sqrt{\lambda_{3}} - \sqrt{\lambda_{4}}\rbrace,
\label{concurrence}
\end{align}
where $\lambda_1 \leqslant  \lambda_2 \leqslant  \lambda_3 \leqslant  \lambda_4$ are the eigenvalues of the matrix $\varrho_{AB} \widetilde{\varrho}_{AB}$, in which $\varrho_{AB}$ is the reduced density matrix for the pair labeled by $A$ and $B$, and 
$\widetilde{\varrho}_{AB}=\sigma_{A}^y \otimes \sigma_{B}^y \varrho_{AB}^{*} \sigma_{A}^y \otimes \sigma_{B}^y$. Note that 
the reduced density matrix can be written in terms of two-point correlation functions as
\begin{align}
\varrho_{AB}=\frac{1}{4}\sum_{i,j=0}^3 \langle \sigma _{A}^{i} \otimes \sigma _{B}^{j} ~\rangle_{AB} \sigma _{A}^{i}\otimes \sigma _{B}^{j},
\label{densitymatrix}
\end{align}
where $(\sigma_0,\sigma_1,\sigma_2,\sigma_3)\equiv (\openone,\sigma_x,\sigma_y,\sigma_z)$. 

Now we aim to calculate concurrence for the KN model with the objective of finding potential quantum critical point(s) and its (their) dependence on the anisotropy parameters. The factor $(1/2)\sigma_{A (B)}$ in Eq.~(\ref{densitymatrix}) can be either $s$ or $\tau$ spins defined in the KN Hamiltonian. We define the following generating functions to calculate the correlation functions from the GSE (obtained in the previous section): 
\begin{align}
\tau _{i} ^{\beta}\tau_{j} ^{\beta} = \frac{1}{d \lambda L^{d}} \frac{\partial H}{\partial \eta_{ \beta}}, \\
\tau _{i} ^{z}s_{i} ^{z}= \frac{1}{d \lambda L^{d}} \frac{\partial H}{\partial \alpha},
\label{generating-functions}
\end{align}
hence we obtain
\begin{align}
\langle \tau _{i} ^{\beta}\tau_{j} ^{\beta} \rangle=
\frac{1}{d \lambda L^{d}} \frac{\partial E_0}{\partial \eta_{\beta}}, \\
\langle \tau _{i} ^{z}s_{i} ^{z} \rangle=
\frac{1}{d \lambda L^{d}} \frac{\partial E_0}{\partial \alpha}.
\label{53}
\end{align}
The explicit expressions of the correlation functions in terms of the model parameters for two neighboring spins are presented in Appendix \ref{correlations}.

If we consider parity \cite{amico2008entanglement}, the definition of concurrence can be simplified as 
\begin{equation}
C=2 \max[0,C^{\mathrm{I}},C^{\mathrm{II}}],
\label{c1}
\end{equation}
where
\begin{align}
C^{\mathrm{I}}&=\vert \langle \tau _{i} ^{x}\tau_{j} ^{x} \rangle+\langle \tau _{i} ^{y}\tau_{j} ^{y} \rangle \vert -\sqrt{(\frac{1}{4}+\langle \tau _{i} ^{z}\tau_{j} ^{z} \rangle)^{2}-m_{z}^{2}}, \label{c2}\\
C^{\mathrm{II}}&=\vert \langle \tau _{i} ^{x}\tau_{j} ^{x} \rangle-\langle \tau _{i} ^{y}\tau_{j} ^{y} \rangle \vert +\langle \tau _{i} ^{z}\tau_{j} ^{z} \rangle)^{2}-\frac{1}{4},\label{c3}
\end{align}
 with $m^{z}$ is the staggered magnetization in the antiferromagnetic case.

\begin{table}[tp]
\caption{The quantum critical point for different values of the anisotropy parameter $\Delta$ in the two-dimensional KN model. 
The (gap closure) PCUT results are compared with the corresponding results from the mean-field~\cite{Langari06} (MF) and Green's function~\cite{rezania} (GF) approaches.}
\begin{tabular}{|c|c|c|c|c|c|}
\hline \hline
\,\,\,\,\, $\Delta$ \,\,\,\,\, & \,\,\,\,\, $0.2$ \,\,\,\,\, & \,\,\,\,\, $0.4$ \,\,\,\,\, & \,\,\,\,\, $0.6$ \,\,\,\,\, & \,\,\,\,\, $0.8$ \,\,\,\,\, & \,\,\,\,\, $1.0$ \,\,\,\,\, \\
\hline $\lambda_c$ (PCUT) & $0.2$ & $0.40$ & $0.56$ & $0.68$ & $0.77$ \\
\hline $\lambda_c$ (MF) & $0.42$ & $0.50$ &  $0.56$ & $0.63$ & $0.70$ \\
\hline $\lambda_c$ (GF) & $0.39$ & $0.46$ & $0.52$ & $0.58$ & $0.65$ \\
\hline
\end{tabular}
\label{qcp2d}
\end{table}

We consider two specific cases here. In the first case, the ground state is supposed to have the $Z_2$ symmetry, i.e., the case of 
no spontaneous symmetry breaking. We find zero magnetization for this ground state, that is, $m^{\alpha}\equiv\langle \tau^{\alpha} \rangle=0$. Thus, concurrence between two neighboring spins of itinerant electrons on the hypercubic lattice is given by the following equation:
\begin{align}
C=&2 \max\Big[0,\Big\vert \sum_{\beta=x, y, z}\langle \tau _{i} ^{\beta}\tau_{i+1} ^{\beta} \rangle \Big\vert -\frac{1}{4}\Big]\nonumber\\
&=2 \max\Big[0,\frac{1}{4} (\eta_{x} +\eta_{y} +\eta_{z}) \overline{\lambda}+ \frac{3}{8}(\eta_{x} \eta_{y}+ \eta_{x} \eta_{z} + \eta_{y} \eta_{z} ) \overline{\lambda} ^{2} \nonumber\\
&~+\frac{\overline{\lambda }^{3}}{8} \big[ (4d-1) (\eta_{x} +\eta_{y} +\eta_{z})(\eta_{x}^{2} +\eta_{y}^{2} +\eta_{z}^{2})  \nonumber\\
& ~+2 ( \eta_{x} \eta_{y} ^{2}+\eta_{y} \eta_{x} ^{2}+\eta_{x} \eta_{z} ^{2}+\eta_{z} \eta_{x} ^{2}+\eta_{y} \eta_{z} ^{2}+\eta_{z} \eta_{y} ^{2} )\nonumber\\
&~+4(4d-3)(\eta_{x}^{3} +\eta_{y}^{3} +\eta_{z}^{3}) \big] -\frac{1}{4}\Big].
\label{c-without-ssb}
\end{align}
From this equation it can be seen for $\lambda<1$, concurrence vanishes for one-, two-, and three-dimensional lattices. This shows that concurrence is not a conclusive tool for presenting QPT in the KN model, which supports previous results for a one-dimensional KN-model \cite{mendoza}. In other words, the correlations responsible for QPT of the KN model are not of bipartite nature captured by  concurrence. The reason should be related to the presence of multipartite entanglements, which cannot be observed by concurrence.   

\section{Summary and discussions}
\label{conclusion}

\begin{table}[tp]
\caption{The GSE per site at different orders of PCUT for the one-dimensional KN model with $(\eta_x, \eta_y, \eta_z)=(1, 1, 1)$ and $\Delta=1$. The relative correction is defined by $|[\epsilon^{(n+1)}-\epsilon^{(n)}]/ \epsilon^{(n)}|$.}
\begin{tabular}{|c|c|c|c|c|c|c|}
\hline\hline	
 $O(\lambda)$ & $\epsilon^{(n)}$ & relative correction\\ \hline
\hline $n=2$  & $-0.421875000$ & $ $\\
\hline $n=3$  & $-0.433593750$ & $0.02777778$\\
\hline $n=4$  & $-0.432861328$ & $0.00168919$\\
\hline $n=5$  & $-0.430114746$ & $0.00634518$\\
\hline $n=6$  & $-0.428932190$ & $0.00274940$\\
\hline $n=7$  & $-0.429188808$ & $0.00059827$\\
\hline $n=8$  & $-0.429747449$ & $0.00130162$\\
\hline\hline
\end{tabular}
\label{table:comp}
\end{table}

We applied the perturbative continuous unitary transformation (PCUT) to study quantum phase transitions and anisotropy effects in the Kondo-necklace model. The advantage of PCUT is that it gives the results for the thermodynamic limit, hence it evades finite-size effects. However, it is essentially a perturbative approach and the accuracy of its results depends on the order of perturbation used to calculate the effective Hamiltonian. We thus obtained the effective Hamiltonian and thereby the ground-state 
energy to the $4$th order of perturbation. A systematic calculation of higher orders of perturbation is also possible.

We also have calculated the ground-state energy up to $O(\lambda^4)$ for the hypercubic Kondo-necklace model in one, two, and three dimensions as well as for the $n$-leg ladder. To justify the accuracy of the PCUT results, we have compared the ground state energy per site for the one-dimensional model with the Lanczos results---see Table \ref{comparison}. The Lanczos computations performed for a chain up to $24$ spins. The finite-size scaling analysis on the Lanczos results shows only a small finite-size effect, which implies that the results for $2L=24$ already exhibited a good approximation of the expected values in the thermodynamic limit. 

Additionally, in order to justify the accuracy of the PCUT results, we have been performed numerical computations \cite{knetter:EJB00,knetter:JPA03} for the ground state energy of the one-dimensional Kondo-necklace model with anisotropic parameters $(\eta_x,\eta_y,\eta_z)=(1,1,1)$ and $\Delta=1$ up to the $8$th order of perturbation. We have calculated the relative correction of successive orders of perturbation $|[\epsilon^{(n+1)}-\epsilon^{(n)}]/ \epsilon^{(n)}|$ (see Table \ref{table:comp}), which demonstrated the accuracy of the perturbative calculations for $\lambda=1$. It has to be mentioned that $\lambda=1$ shows the worst case in the perturbation scheme and we would expect better accuracy for smaller parameter ($\lambda <1$). Higher orders of perturbation improve the accuracy of the ground-state energy, however we get three  digits of accuracy if we keep the $4$th order correction in the ground-state energy. Hence, we conclude that the calculation up to $O(\lambda^4)$ gives reliable data for our analysis, which has been taken into account in this  article. Moreover, the $4$th order correction is given in a closed analytical form, while higher orders can only be obtained numerically. The analytic form has the benefit to define generating functions in terms of derivatives of the couplings, which would lead to the calculation of correlation functions and concurrence.

\begin{figure}[tp]
\includegraphics[width=8cm]{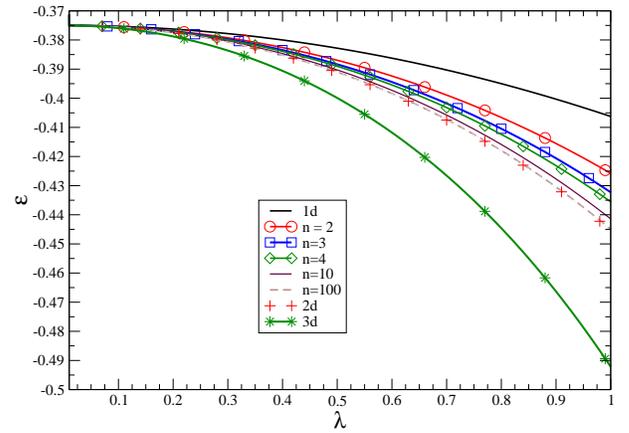}
\caption{The GSE per site for the one-, two-, three-dimensional hypercubic KN model, and the $n$-leg ladder.}
\label{energy-ladder}
\end{figure}

\begin{table}[bp]
\caption{Quantum critical points ($\lambda_c$) of the isotropic two- and three-dimensional KN model obtained by the mean-field~\cite{Langari06} (MF), Green's function~\cite{rezania} (GF), and zero gap through PCUT ($G_2=0$ or $G_3=0$).}
\begin{tabular}{|c|c|c|c|}
\hline\hline	
$\lambda_c$ & MF & GF  & PCUT
\\ \hline
\hline  $d=2$ &0.70 & 0.65 & 0.77 \\
\hline  $d=3$ &0.40 & -- & 0.59 \\
\hline
\end{tabular}
\label{qcp-2d-3d}
\end{table}

The excitation energies of the Kondo-necklace model and the dispersion relations were obtained analytically within the PCUT
formalism. The minimum of the excitations defines the energy gap, which vanishes at the quantum critical point. The quantum critical points of the two- and three-dimensional Kondo-necklace model and their dependence on the anisotropy parameter $\Delta$ were presented in Table \ref{qcp2d} and Figs.~\ref{gap2d} and \ref{gap3d}. Both figures showed that the associated criticality depends relatively strongly on $\Delta$. More information on the nature of quantum phase transition(s) in the Kondo-necklace model can be obtained by computing entanglement in the system. Thus, we calculated concurrence without spontaneous symmetry breaking (a measure of bipartite entanglement) in the PCUT formalism through calculating generating functions of two-point correlation functions. For $\lambda < 1$, we always obtained zero for the value of the concurrence. This implies concurrence is not sensitive to the quantum phase transition in Kondo-necklace model in agreement with the results reported for one-dimensional Kondo-necklace model \cite{mendoza}. This can be interpreted as the presence of multipartite entanglement close to the quantum critical point which can be deduced with other measures of entanglement i.e. von Neumann block entropy \cite{mendoza}. We have also studied concurrence in the presence of spontaneous symmetry breaking---as suggested in Ref. \onlinecite{oliveira2008} for the generalized entanglement entropy---which did not change our results, i.e., zero value of concurrence for $\lambda <1$. We have imposed spontaneous symmetry breaking explicitly via a mean-field value for the staggered magnetization that has not been shown here.

The calculated values of the quantum critical point $\lambda_c$ for the two- and three-dimensional models are less than one. In fact, $\lambda_c\ll 1$ ($\lambda_c < 1$) justifies the accuracy of PCUT. We should note that away from the critical point the perturbation series converges even at low orders of perturbation, but it hardly converges close to the critical point. Despite this shortcoming of perturbation theory close to critical point, the results for critical points summarized in Table \ref{qcp-2d-3d} are in good agreements with other approaches. The mean-field approach for $d=1$ \cite{Langari06} yields a Kondo-singlet phase for the whole range of $\lambda$, hence, no quantum phase transition. One can argue based on a general qualitative description no quantum phase transition occurs for the $U(1)$-symmetric one-dimensional Kondo-necklace model because no long-range order for both extreme limits of $\lambda$ is seen: $\lambda=0$ gives the Kondo-singlet state and $\lambda \rightarrow \infty$ leads to the one-dimensional spin-$1/2$ XXZ model, which has a spin-fluid ground state \cite{sachdev:book} with no long-range order. Thus the very existence of a quantum phase transition in the one-dimensional model is not justified; this still requires more exhaustive analysis. In contrast, the $Z_2$-symmetric Kondo-necklace model shows a quantum phase transition from the Kondo-singlet phase to the antiferromagnetic ordered phase \cite{mahmoudian}.

A particularly interesting result of our formalism is an observation about the $n$-leg ladder. The plot of the ground-state energy 
per spin ($\varepsilon$) for the $n$-leg ladder in addition to the one-, two-, and three-dimensional models in Fig. \ref{energy-ladder} indicates that for the ladder the values of $\varepsilon$ are visibly more similar to its value for $d=2$ rather than for $d=1$. Clearly, the ground-state energy per spin reaches the two-dimensional values in the $n\rightarrow \infty$ limit. It can be understood from Eqs.~\eqref{gse-perspin} and \eqref{ladder-energy} in which some extra terms of the ladder ground-state energy will be omitted for $n=1$. Moreover, similar to the two-dimensional model, the concurrence of ladder without spontaneous symmetry breaking does not show a singular derivative, manifesting that bipartite entanglement does not indicate a possible quantum phase transition. We thus conclude that the $n$-leg Kondo-necklace model exhibits a quantum phase transition from the Kondo-singlet phase to the antiferromagnetic-ordered state, similarly to the two-dimensional Kondo-necklace model. However, we caution that the perturbative nature of PCUT and the $4$th order calculations implemented here could lead to artifacts. Thus more careful investigation on the delicate question of crossover from one to two spatial dimensions is still necessary. \cite{mmtalk2015}

\textit{Acknowledgments.---}We thank S. Mahmoudian for fruitful discussions. This work was supported in part by Sharif University of Technology's Center of Excellence in Complex Systems and Condensed Matter and the Office of Vice President for Research. A. L. acknowledges partial support from the Alexander von Humboldt Foundation.



\newpage
\begin{widetext}
\appendix

\section{Effect of the $q_n$ operators}
\label{aa}

\begin{table}[!h]
\caption{Effect of $q_{n}$ on unperturbed bipartite eigenstates. Here $|s \rangle$ represents a singlet state on a dimer, while $
|t^{\nu} \rangle$ represents a triplet on a dimer with $\nu=0, \pm 1$ as its total $z$-component spin.}
\begin{tabular}{|c c c|}
\hline\hline
 & $4 q_{0}$ & \\
\hline
$| s,s \rangle$ & $\longrightarrow$ & $-2 \alpha | s,s \rangle$\\
$| t^{\pm 1},s \rangle$ & $\longrightarrow$ & $\frac{1}{2} (\eta_{x}+\eta_{y}) | s , t^{\pm 1} \rangle  - \frac{1}{2} (\eta_{x}-\eta_{y}) | s , t^{\mp 1} \rangle$ \\
$| s,t^{\pm 1} \rangle$ & $\longrightarrow$ & $\frac{1}{2} (\eta_{x}+\eta_{y}) | t^{\pm 1},s \rangle - \frac{1}{2} (\eta_{x}-\eta_{y}) | t^{\mp 1},s \rangle $\\
$| t^{0},s \rangle$ & $\longrightarrow$ & $ -2 \alpha | t^{0}, s \rangle + \eta_{z} | s,t^{0} \rangle$\\
$| s,t^{0} \rangle$ & $\longrightarrow$ & $ -2 \alpha | s , t^{0} \rangle + \eta_{z} | t^{0},s \rangle$ \\
$| t^{0},t^{0} \rangle$ & $\longrightarrow$ & $-2 \alpha | t^{0},t^{0}\rangle +\frac{1}{2} (\eta_{x}+\eta_{y}) (| t^{1},t^{-1} \rangle + | t^{-1},t^{1} \rangle) +\frac{1}{2} (\eta_{x}-\eta_{y}) (| t^{1},t^{1} \rangle + | t^{-1},t^{-1} \rangle) $ \\
$| t^{\pm 1},t^{\pm 1} \rangle$ & $\longrightarrow$ & $(2 \alpha+\eta_{z})| t^{\pm 1},t^{\pm 1} \rangle + \frac{1}{2} (\eta_{x}-\eta_{y}) | t^{0},t^{0} \rangle $ \\
$| t^{\pm 1},t^{\mp 1} \rangle$ & $\longrightarrow$ & $(2 \alpha-\eta_{z})| t^{\pm 1},t^{\mp 1} \rangle + \frac{1}{2} (\eta_{x}+\eta_{y}) | t^{0},t^{0} \rangle $ \\
$| t^{\pm 1},t^{0} \rangle$ & $\longrightarrow$ &  $\frac{1}{2} (\eta_{x}+\eta_{y}) | t^{0},t^{\pm 1} \rangle + \frac{1}{2} (\eta_{x}-\eta_{y}) | t^{0},t^{\mp 1} \rangle$ \\
$| t^{0},t^{\pm 1} \rangle$ & $\longrightarrow$ &  $\frac{1}{2} (\eta_{x}+\eta_{y}) | t^{\pm 1},t^{0} \rangle+ \frac{1}{2} (\eta_{x}-\eta_{y}) | t^{\mp 1},t^{0} \rangle$ \\
\hline
& $4 q_{1}$ & \\
\hline
$| t^{\pm 1},s \rangle $ & $\longrightarrow$ & $ \pm \eta_{z} | t^{\pm 1},t^{0} \rangle \mp \frac{1}{2} (\eta_{x}+ \eta_{y}) | t^{0},t^{\pm 1} \rangle \pm \frac{1}{2} (\eta_{x} - \eta_{y}) | t^{0},t^{\mp 1} \rangle$\\
$| s,t^{\pm 1} \rangle $ & $\longrightarrow$ & $ \pm \eta_{z} | t^{0},t^{\pm 1} \rangle \mp \frac{1}{2} (\eta_{x}+ \eta_{y}) | t^{\pm 1},t^{0} \rangle \pm \frac{1}{2} (\eta_{x} - \eta_{y}) | t^{\mp 1},t^{0} \rangle$\\
 $| t^{0},s \rangle $ & $\longrightarrow$ & $ \frac{1}{2} (\eta_{x}+ \eta_{y}) (| t^{1},t^{-1} \rangle - | t^{-1},t^{1} \rangle)+\frac{1}{2} (\eta_{x} - \eta_{y}) (| t^{-1},t^{-1} \rangle - | t^{1},t^{1} \rangle) $\\
$| s,t^{0} \rangle $ & $\longrightarrow$ & $ \frac{-1}{2} (\eta_{x}+ \eta_{y}) (| t^{1},t^{-1} \rangle - | t^{-1},t^{1} \rangle)+\frac{1}{2} (\eta_{x} - \eta_{y}) (| t^{-1},t^{-1} \rangle - | t^{1},t^{1} \rangle) $ \\
\hline
& $4 q_{2}$ & \\
\hline
 $| s,s \rangle$ & $\longrightarrow$ & $\eta_{z} | t^{0},t^{0} \rangle -\frac{1}{2} (\eta_{x}+ \eta_{y})(| t^{1},t^{-1} \rangle + | t^{-1},t^{1} \rangle )+\frac{1}{2} (\eta_{x} - \eta_{y}) (| t^{-1},t^{-1} \rangle + | t^{1},t^{1} \rangle ) $ \\
\hline
& $4 q_{-1}$ & \\
\hline
$| t^{\pm 1},t^{\pm 1} \rangle $  & $\longrightarrow$ & $\mp \frac{1}{2} (\eta_{x}- \eta_{y})( | t^{0},s \rangle + | s,t^{0} \rangle ) $\\
$| t^{\pm 1},t^{\mp 1} \rangle $  & $\longrightarrow$ & $\pm \frac{1}{2} (\eta_{x}- \eta_{y})( | t^{0},s \rangle - | s,t^{0} \rangle )$ \\
$| t^{\pm 1},t^{0} \rangle $ & $\longrightarrow$ & $\pm \eta_{z} | t^{\pm1},s \rangle \mp \frac{1}{2} (\eta_{x}+ \eta_{y}) | s,t^{\pm 1} \rangle \mp \frac{1}{2} (\eta_{x} - \eta_{y}) | s,t^{\mp 1} \rangle  $ \\
$| t^{0},t^{\pm 1}, \rangle $ & $\longrightarrow$ & $\pm \eta_{z} | s,t^{\pm1} \rangle \mp \frac{1}{2} (\eta_{x}+ \eta_{y}) | t^{\pm 1},s \rangle \mp \frac{1}{2} (\eta_{x} - \eta_{y}) | ,t^{\mp 1} s\rangle  $ \\
\hline
& $4 q_{-2}$ & \\
\hline
$| t^{\pm 1},t^{\pm 1} \rangle $  & $\longrightarrow$ & $\frac{1}{2} (\eta_{x} - \eta_{y}) | s,s \rangle $\\
$| t^{0},t^{0} \rangle $ & $\longrightarrow$ & $\eta_{z} | s,s \rangle $\\
$| t^{\pm 1},t^{\mp 1} \rangle $  & $\longrightarrow$ & $-\frac{1}{2} (\eta_{x} + \eta_{y}) | s,s \rangle $\\
\hline
\hline
\end{tabular}
\label{jadv}
\end{table}

\section{Spectrum of two- and three-dimensional model}
\label{spectrum}

The excitation spectrum of the two-dimensional model is given by the following equation: 
\begin{align}
\label{2d-dispersion}
\omega(k)=&1+\overline{\lambda}\Big(2[\cos(k_{x})+\cos(k_{y})]+8\alpha\Big)
-\overline{\lambda}^{2}\Big(-4\eta_z[\cos(k_{x})+\cos(k_{y})]+\cos(2k_{x})
+2[\cos(k_{x}+k_{y})+\cos(k_{x}-k_{y})]\nonumber \\
&+\cos(2k_{y})+2\eta_z^{2}\Big)
+\overline{\lambda}^{3}\Big(-\frac{21\eta_z^{2}+23}{2}[\cos(k_{x})+\cos(k_{y})]
+3[\cos(k_{x}+2k_{y})+\cos(k_{x}-2k_{y})+\cos(2k_{x}+k_{y}) \nonumber \\
&+\cos(2k_{x}-k_{y})]+\cos(3k_{x})+\cos(3k_{y})
+(16\alpha +8\eta_z)[\cos(k_{x}+k_{y})+\cos(k_{x}-k_{y})]+16\eta_z+32\alpha -10\Big).
\end{align}
where ($k_x, k_y$) represents the momentum components.

The one-magnon spectrum of the three-dimensional model is given by the following expression:
 \begin{equation}
\label{3d-dispersion}
\omega(k)=1+\overline{\lambda} [2 a_1+12\alpha]-\overline{\lambda}^{2}[3 {\eta_z}^2-4\eta_z a_1 +2a_3 + a_2]+\overline{\lambda}^{3} [-72\alpha+(12\alpha + 4\eta_z)a_2
+(24\alpha+8\eta_z)a_3-\frac{5}{2}a_{1}(9+7\eta_z^{2})+9\eta_z],
\end{equation}
in which
\begin{align}
& a_1 = \cos(k_{x}) + \cos(k_{y}) +\cos(k_{z}),  \nonumber \\
& a_2 = \cos(3k_{x}) + \cos(3k_{y}) +\cos(3k_{z}), \nonumber\\
& a_3 = \cos(k_{x}+k_{y})+\cos(k_{x}-k_{y})+\cos(k_{x}+k_{z})+\cos(k_{x}-k_{z})+\cos(k_{y}+k_{z})+\cos(k_{y}-k_{z}).   \nonumber
\end{align}

\section{Correlation functions}
 \label{correlations}

The correlation function of the nearest-neighbor spins on the $d$-dimensional hypercubic KN lattice can be obtained via the generating functions defined in Eq.~\eqref{generating-functions}. The correlations functions have simple forms for the isotropic case in the local interactions $\alpha=0$ ($\Delta=1$) 
\begin{align}
\langle \tau _{i} ^{x}\tau _{i+1} ^{x} \rangle= -\frac{1}{4} \eta_{x} \overline{\lambda}- \frac{3}{8} \eta_{y} \eta_{z} \overline{\lambda} ^{2}+ \frac{\overline{\lambda }^{3}}{32} [- 8 \eta_{x}( \eta_{y} ^{2}+\eta_{z} ^{2}) +4(4d-1) \eta_{x} (\eta_{x}^{2} +\eta_{y}^{2} +\eta_{z}^{2})-16(4d-3)\eta_{x}^{3} ], \label{55} \\
\langle \tau _{i} ^{y}\tau _{i+1} ^{y} \rangle=- \frac{1}{4} \eta_{y} \overline{\lambda}
- \frac{3}{8} \eta_{x} \eta_{z} \overline{\lambda} ^{2}
+ \frac{\overline{\lambda }^{3}}{32} [ - 8 \eta_{y}( \eta_{x} ^{2}+\eta_{z} ^{2}) +4 (4d-1)\eta_{y} (\eta_{x}^{2} +\eta_{y}^{2} +\eta_{z}^{2})- 16(4d-3)\eta_{y}^{3} ], \label{56}\\
\langle \tau _{i} ^{z}\tau _{i+1} ^{z} \rangle=- \frac{1}{4} \eta_{x} \overline{\lambda}
-\frac{3}{8} \eta_{x} \eta_{y} \overline{\lambda} ^{2}
+\frac{\overline{\lambda }^{3}}{32} [- 8 \eta_{z}( \eta_{x} ^{2}+\eta_{y} ^{2}) +4 (4d-1) \eta_{z} (\eta_{x}^{2} +\eta_{y}^{2} +\eta_{z}^{2}) -16(4d-3)\eta_{z}^{3} ]. \label{57}
\end{align}

The analytical expression for the local correlation functions between the impurity spin ($s$) and the spin on the lattice ($\tau$) are in the following form:
\begin{align}
\langle \tau _{i} ^{x} s _{i} ^{x} \rangle=- \frac{1}{4}
+ \frac{d}{4}(\eta_{y}^{2}+\eta_{z}^{2}) \overline{\lambda} ^{2}
-d \overline{\lambda }^{3} \eta_{x}\eta_{y}\eta_{z},  \\
\langle \tau _{i} ^{y} s _{i} ^{y} \rangle=- \frac{1}{4}
+ \frac{d}{4}(\eta_{x}^{2}+\eta_{z}^{2}) \overline{\lambda} ^{2}
-d \overline{\lambda }^{3} \eta_{x}\eta_{y}\eta_{z},  \\
\langle \tau _{i} ^{z} s _{i} ^{z} \rangle=
-\frac{1}{4} + \frac{d}{4}(\eta_{x}^{2}+\eta_{y}^{2}) \overline{\lambda} ^{2}
-d \overline{\lambda }^{3} \eta_{x}\eta_{y}\eta_{z}.
\label{550}
\end{align}

The correlation functions for the $n$-leg ladder are as follows:
\begin{align}
\langle \tau^x_{1,i} \tau^x_{1,i+1} \rangle
=& -\frac{\overline{\lambda}} {4}\eta_{x}-\frac{3\overline{\lambda}^{2}} {8}\eta_{y} \eta_{z} +\frac{\overline{\lambda}^{3}}{8}[3\eta_{x} ( \eta_{x} ^{2}+  \eta_{y} ^{2}+  \eta_{z} ^{2} )
+4\eta_{x} x^{2} \frac{(n-1)}{n}( \eta ^{'2}_{x}+  \eta ^{'2}_{y}+  \eta^{'2}_{z}) -2\eta_{x}(\eta_{y}^{2}+\eta_{z}^{2}+2\eta_{x}^{2}) \nonumber\\ &-16\frac{(n-1)}{n}x^{2}(\eta_{x})(\eta ^{'2}_{x})],\\
\langle \tau^y_{1,i} \tau^y_{1,i+1} \rangle
= & -\frac{\overline{\lambda}} {4}\eta_{y}-\frac{3\overline{\lambda}^{2}} {8}\eta_{x} \eta_{z} +\frac{\overline{\lambda}^{3}}{32}[3\eta_{y} ( \eta_{x} ^{2}+  \eta_{y} ^{2}+  \eta_{z} ^{2} )+4\eta_{y} x^{2} \frac{(n-1)}{n}( \eta ^{'2}_{x}+  \eta ^{'2}_{y}+  \eta^{'2}_{z}) -2\eta_{y}(\eta_{x}^{2}+\eta_{z}^{2}+2\eta_{y}^{2}) \nonumber\\ 
&-16\frac{(n-1)}{n}x^{2}(\eta_{y})(\eta ^{'2}_{y})],\\
\langle \tau^z_{1,i} \tau^z_{1,i+1} \rangle
= & -\frac{\overline{\lambda}} {4}\eta_{z}-\frac{3\overline{\lambda}^{2}} {8}\eta_{x} \eta_{y}+\frac{\overline{\lambda}^{3}}{8}[3\eta_{z} ( \eta_{x} ^{2}+  \eta_{y} ^{2}+  \eta_{z} ^{2} )+4\eta_{z} x^{2} \frac{(n-1)}{n}( \eta ^{'2}_{x}+  \eta ^{'2}_{y}+  \eta^{'2}_{z}) -2\eta_{z}(\eta_{x}^{2}+\eta_{y}^{2}+2\eta_{z}^{2}) \nonumber\\ 
&-16\frac{(n-1)}{n}x^{2}(\eta_{z})(\eta ^{'2}_{z})], \\
\langle \tau^x_{1,i} \tau^x_{2,i} \rangle
= & -\frac{\overline{\lambda}} {4}\eta^{'}_{x}-\frac{3 x^{3}\overline{\lambda}^{2}} {8}\eta^{'}_{y} \eta^{'}_{z} +\frac{\overline{\lambda}^{3}}{8}[4x^{2}\eta^{'}_{x} ( \eta_{x} ^{2}+  \eta_{y} ^{2}+  \eta_{z} ^{2} )+3x^{4}\eta^{'}_{x}( \eta ^{'2}_{x}+  \eta ^{'2}_{y}+  \eta^{'2}_{z}) \nonumber\\ 
&-16 x^{2}(\eta^{'}_{x}\eta _{x}^{2})-2x^{4}\eta^{'}_{x}(\eta^{'2}_{y}+\eta^{'2}_{z})-4x^{4}\eta^{'3}_{x}], \\
\langle \tau^y_{1,i} \tau^y_{2,i} \rangle
= & -\frac{\overline{\lambda}} {4}\eta^{'}_{y}-\frac{3 x^{3}\overline{\lambda}^{2}} {8}\eta^{'}_{x} \eta^{'}_{z}+\frac{\overline{\lambda}^{3}}{8}[4x^{2}\eta^{'}_{y} ( \eta_{x} ^{2}+  \eta_{y} ^{2}+  \eta_{z} ^{2} )+3x^{4}\eta^{'}_{y} ( \eta ^{'2}_{x}+  \eta ^{'2}_{y}+  \eta^{'2}_{z} ) \nonumber\\
&-16 x^{2}(\eta^{'}_{y}\eta _{y}^{2})-2x^{4}\eta^{'}_{y}(\eta^{'2}_{x}+\eta^{'2}_{z})-4x^{4}\eta^{'3}_{y}],\\
\langle \tau^z_{1,i} \tau^z_{2,i} \rangle
= & -\frac{\overline{\lambda}} {4}\eta^{'}_{z}-\frac{3 x^{3}\overline{\lambda}^{2}} {8}\eta^{'}_{x} \eta^{'}_{y} +\frac{\overline{\lambda}^{3}}{8}[4x^{2}\eta^{'}_{z} ( \eta_{x} ^{2}+  \eta_{y} ^{2}+  \eta_{z} ^{2} )+3x^{4}\eta^{'}_{z} ( \eta ^{'2}_{x} +  \eta ^{'2}_{y}+  \eta^{'2}_{z}) \nonumber\\ 
& -16x^{2}(\eta^{'}_{z}\eta _{z}^{2})-2x^{4}\eta^{'}_{z}(\eta^{'2}_{x}+\eta^{'2}_{y})-4x^{4}\eta ^{'3}_{z}].
\end{align}
\twocolumngrid
\end{widetext}

\end{document}